# Between Algorithm (AI) and Intuition (Human): Preserving Designer Agency in AI-Assisted Sensemaking of Qualitative UX Data

Md Haseen Akhtar, Department of Design, IIT Hyderabad, haseen@des.iith.ac.in

## ABSTRACT

*The integration of AI into qualitative design research presents a fundamental tension: how do we leverage AI while preserving the subjective, intuitive judgments that define design expertise? This paper examines this question through a case study of analyzing 20 user responses about video conferencing platforms for educational contexts. We argue that AI sensemaking tools risk flattening the rich data patterns, amplifying contradictory textures of user feedback into sterile categories thereby transforming design research from an interpretive craft into a mechanical sorting exercise (rigid and formal). Through comparative analysis of AI-assisted sensemaking versus human-centered approaches to the same dataset, we identify when algorithmic efficiency enhances understanding and when it diminishes the designer's interpretive agency (uncovering hidden needs, critical enquiry, what if enquiries, making decisions, having trade-offs). We present a framework for augmented sensemaking that positions AI as an instrument for amplifying human judgment rather than replacing it. Our findings suggest that the most valuable role for AI in design research is not to eliminate subjectivity, but to make it more intentional, reflective, and accountable.*

## INTRODUCTION

Design research is inherently interpretive. When students were asked about engagement tracking features for instructors to use, they said *"I think just knowing who's active and how much is enough,"*. If a designer tries to find the latent needs from this statement, it could mean multiple things. Are they uncomfortable with being watched but accepting it anyway? Are they saying, this much tracking is okay, but no more? or are they worried about surveillance but not saying so directly? AI can label this response as engagement tracking preferences, but it cannot detect the uncertainty in their words or understand the deeper concerns they might reveal, if asked follow-up questions [1].

This tension between AI and interpretive depth animates our research question: **Can AI assist sensemaking without suppressing the designer's interpretive agency?** The promise of AI in qualitative analysis is seductive: faster thematic coding, pattern recognition across large datasets, reduction of cognitive load [2]. Yet design is fundamentally subjective, it privileges certain user voices, reads between lines, makes aesthetic and ethical judgments that resist formalization [3]. Recent advances in large language models have made AI-assisted qualitative analysis increasingly accessible to design practitioners. AI tools like Claude, ChatGPT-5, Gemini, Perplexity, Grok, Llama and Deepseek V3.2 promise to transform how designers make sense of user data [4]. However, the integration of these tools raises critical questions about the nature of design knowledge itself. If an AI can identify themes, extract quotes, and generate summaries, what remains distinctively human about the sensemaking process?

## BACKGROUND: SENSEMAKING AS INTERPRETIVE LABOR

Sensemaking, as theorized by Russell et al. [1], involves the iterative process of searching for representations and encoding data into these representations to answer questions. In design contexts, this process is deeply interpretive, designers do not simply discover patterns in data, they construct them through acts of selection, emphasis, and framing [5]. Hook and Lowgren [3] argue that design research produces strong concepts, intermediate-level knowledge that bridges theory and practice through situated interpretation. These concepts emerge not from objective analysis but from the designer's ability to recognize significance, make connections, and exercise taste. A quote becomes meaningful not because it appears frequently, but because a designer judges it to be revealing of deeper patterns.

Recent work on data-centric sensemaking emphasizes the embodied, collaborative nature of working with qualitative data [6]. Koesten et al. [7] demonstrate that data sensemaking involves not just cognitive processes but social negotiations about what counts as evidence, what interpretations are valid, and whose perspectives matter. This inherently political dimension of sensemaking, deciding which user needs to prioritize, which contradictions to resolve, cannot be automated without also automating the value judgments embedded in these decisions. *Our argument is that AI tools cannot reveal these deeper hidden meanings in multiple layers since they are far away from the situated reality of the problem space.*

The introduction of AI tools into this process raises questions about the distribution of interpretive authority. When AI categorizes user feedback, whose categories are being applied? When it identifies key themes, by what criteria are some themes deemed more key than others? As Suchman [8] argues, seemingly neutral technical systems always encode worldviews and power relations. Thereby, they cannot make decisions based on value systems for the context in enquiry as it should be in a design activity.

## METHOD

We conducted a structured comparison using Claude 3.5 Sonnet to analyze our dataset. The methodology involved: (1) providing the AI with the complete dataset and research context, (2) requesting thematic analysis with supporting quotes, (3) comparing AI-generated themes with themes identified through traditional close reading and qualitative coding, and (4) examining specific instances where interpretations diverged.

We explore this through a dataset of 20 student responses (comprising approximately 4,500 words across 40 questions) about video conferencing platforms for teaching and learning. By comparing AI-generated analyses with a designer's close reading of the same material, we examine where algorithmic sensemaking elucidates and where it obscures. Our goal is not to reject Gen-AI tools, but to understand their appropriate scope and to develop practices that preserve designer agency (end goal) while leveraging computational capabilities through these AI tools.

## RESULTS AND DISCUSSION

## THE DESIGN SENSEMAKING CHALLENGE

Our dataset reveals the textured complexity typical of qualitative design research. Twenty students responded to questions about their experiences with video conferencing platforms (Zoom, Google Meet, Microsoft Teams) in academic contexts. Their responses contain rich contradictions, unstated assumptions, and affective nuances that resist straightforward categorization.

### Contradictions as Design Material

Students overwhelmingly prefer dark mode (18 of 20 respondents), yet also express concern about eye fatigue from excessive screen time. They want private messaging restricted to reduce distraction but simultaneously value it as an accessibility feature for anxious students who fear speaking aloud. They praise minimal interfaces while requesting more interactive tools, better notifications, and gamification elements. One student noted: *"It should not be only about finishing a task or a topic but also making sure that everyone is comfortable enough and is not getting overloaded with information."* These contradictions are not errors to be resolved through averaging or using majority rule, they are the *substance* of design insight, hidden layers to be revealed and contested upon for making design decisions. They reveal competing values (efficiency versus comfort), tensions between individual and collective needs (personal distraction versus accessibility), and the gap between stated preferences and underlying anxieties.

**Emotional Labor and Uncertainty**

A student who says *"the interface is mostly good but needs some minute tweaks here and there"* is performing emotional labor, softening criticism, concealing uncertainty. An AI analyzing this sentiment might extract positives with minor concerns. A designer reads the subtext: *"I am somewhat frustrated but do not want to seem demanding or complaining about."* This distinction matters because it suggests the student may have more substantial critiques, they are reluctant to voice a hypothesis worth exploring in follow-up research. When discussing private chat restrictions, students exhibit notable uncertainty: *"Students like this feature and say that it is helpful but it kind of contradicts my idea of making the platform more engaging."* The use of "kind of" signals uncertainty, an acknowledgment that the student recognizes conflicting values and social connection through chat versus focused attention on instruction.

**Silences and Absences**

What students do not mention can be as revealing as what they do. Despite extensive discussion of screen sharing as essential (*"meetings can not run smoothly without screen sharing"*), no student expressed privacy concerns about accidentally sharing sensitive content, a common worry in professional contexts [9]. This absence might indicate that students either do not consider their screens private, or they have not yet experienced embarrassing disclosure. Either interpretation has design implications. While students request engagement tracking for instructors, they frame it cautiously: *"Not hindering anyone's privacy should be a priority."* The defensive phrasing (not hindering) suggests anxiety about surveillance on their side and the instructor side as well, that might not be captured by simple keyword analysis of privacy mentions.

## COMPARATIVE SENSEMAKING: AI VERSUS DESIGNER

### What AI does well

The AI excelled at rapid summarization and pattern recognition. Within minutes, it produced clean taxonomies organized around engagement tools, collaboration features, and accessibility concerns. It accurately identified frequently mentioned features (chat: 20 out of 20 responses, screen sharing: 19 out of 20, layout customization: 18 out of 20) and quantified sentiment distributions across questions. For high-volume analysis, this capability is genuinely valuable; it prevents the fatigue-induced errors that are bound to happen during human coding of large datasets.

The AI also surfaced and amplified low frequency suggestions from the users that could be overlooked in manual analysis. Only one student mentioned voice typing as an accessibility feature, but the AI flagged it as a potentially significant feature given the context of reducing distraction and supporting diverse participation styles. This demonstrates AI's utility for ensuring comprehensiveness, making sure rare but valuable insights are not lost.

### Where AI falls short

**1. Flattening Affect and Certainty**

The AI classified *"I think it will make a difference"* and *"It should definitely be a must have"* as both expressing positive sentiment toward agenda-setting features. However, the uncertainty in the first quote (I think) versus the emphatic certainty in the second (definitely, must) reveals different levels of conviction that might inform prioritization. A designer might weigh the second response more heavily or seek to understand why the first student is uncertain.

**2. Missing Situated Enquiry**

As noted earlier, students never mentioned screen sharing privacy concerns, a notable absence. The AI did not flag this gap because it was trained to find what's present, not to notice meaningful silences. A designer familiar with the domain knows that screen sharing anxiety is common [9], and its absence here prompts questions: Have these students not yet encountered this problem? Are they normalizing surveillance? Do they trust their instructors implicitly?

**3. Decontextualizing Institutional Critique**

When one student wrote *"Teachers do not encourage separate group discussions,"* the AI tagged this as a feature request (suggestion: make breakout rooms more prominent). A designer reads it as institutional critique or pedagogical choice, not just technology affordances. This student is observing that the problem is not the technology; it is teaching practices. Addressing this requires curricular intervention, not UI redesign.

**4. Overlooking Self-Questioning and Doubt**

Students frequently questioned whether features would actually be used: *"Do people not know which situation to use and what layout?"* and *"Is this extra work because the normal interface does the job quite well?"* This self-questioning, doubting one's own suggestions, reveals user uncertainty that should shape design priorities. If users are not confident a feature will be adopted, perhaps education and onboarding matter more than additional functionality. The AI coded these as feature concerns rather than recognizing them as epistemological hesitation about user behavior.

**5. Ignoring Power Dynamics**

Our survey asked about engagement indicators for instructors, implicitly framing surveillance as desirable and necessary. Students complied with this framing, suggesting metrics like activity tracking and attention monitoring. However, their language reveals discomfort: *"Not hindering anyone's privacy should be a priority."* The AI did not challenge the premise of the question or note the tension between requesting surveillance features while expressing privacy concerns. A critical designer might question whether the research itself is reproducing problematic assumptions about teacher-student power relations.

## COMPARATIVE SYNTHESIS: WHEN AI HELPS AND WHEN IT HINDERS

Our analysis suggests that AI sensemaking tools are most valuable for **breadth** and **consistency**, but struggle with **depth** and **criticality**. **AI excels at** the following abilities:

**1. Volume and speed:** Capable of processing hundreds of responses without fatigue-induced errors. Example from our data: When analyzing responses for Q11 about tab switching, AI instantly quantified that 95 per cent of students (19/20) report frequently switching tabs during meetings. It cross-referenced this with Q12 responses showing 85 per cent find this distracting, immediately surfacing a high-priority pain point. Manual coding might have taken hours to establish this correlation; AI did it in seconds. For large-scale studies with hundreds of participants, this speed differential becomes critical, preventing research bottlenecks and enabling iterative refinement of interview protocols.

**2. Comprehensiveness:** Does amplification of low frequency user suggestions and ensures no response is overlooked. Example from

our data: Only one student mentioned voice typing as an alternative to text chat (Q2: voice note or voice typing (speech to text feature). AI flagged this low-frequency mention that a fatigued human coder might have overlooked after reading 18 similar responses about delayed messages and missed chats. The suggestion is particularly valuable because it addresses both accessibility (reducing typing burden) and efficiency (faster input). Also, since only one person mentioned it, it could easily be ignored as just one person's odd idea instead of a valuable insight.

**3. Initial organization:** Creating preliminary taxonomies and groupings that can seed human interpretation. Example from Our Data: AI rapidly grouped the 40 questions into logical clusters: engagement mechanisms (Q1-Q13), interaction tools (Q1-Q9 of section B), collaboration features (Q1-Q7 of section C), and meta-level concerns (section D). This preliminary structure revealed that students spent disproportionate time discussing engagement compared to collaboration, a pattern that might inform which areas need deeper investigation. The AI's taxonomy provided scaffolding for human analysis without determining final interpretations.

**4. Quantification:** Counting mentions, tracking sentiment distributions, identifying frequency patterns. Example from Our Data: When students responded to Q9 about color themes, AI quickly established that 75 per cent prefer dark mode, while tracking that their justifications varied significantly with some citing eye strain reduction, others emphasizing aesthetic preference, still others noting improved focus. This mixed quantitative-qualitative summary helps designers understand both the what (strong preference) and the why (diverse motivations), informing whether to make dark mode default or customizable.

**AI Struggles with:**

**1. Interpretation in layers:** Gen-AI cannot read subtext, recognizing uncertainty and emotional labor, understanding what users mean versus what they say. In Q5 responses about layout customization, one student wrote: *"It is definitely useful but in my opinion, people use it more for comfort rather than there being a strong need to use it (except for cases like presentations)."* AI labeled this as positive feedback about the feature. But a designer sees deeper meanings such as the student is unsure, they separate comfortable from necessary (meaning it is not essential), and they mention exceptions (showing it only helps in certain situations). The student is really saying the feature is a could have category but not must have category, which is the opposite of useful when deciding what to build.

**2. Noticing absences:** Despite talking extensively about screen sharing (Q1-Q3 in section C), not one student mentioned worrying about accidentally sharing private content like browser history, personal messages, or financial information. This is surprising because research on remote work shows that screen sharing anxiety is very common. This silence suggests two possibilities: either students do not think of their screens as private (which is worrying for their digital awareness), or they simply have not experienced an embarrassing mistake yet (meaning we need to design protections before problems happen, not after). AI did not notice this gap because it only looks for what people actually say, not what's notably absent.

**3. Critical framing:** Challenging research premises, questioning whose interests are served by designs. Example from our data: Our survey design itself encodes assumptions. Section D, Q3 asks hosts, *"In what ways meeting platforms can include tracking of participants: engagement or attentiveness?"* The question presumes surveillance is legitimate and desirable. Students complied with this framing, suggesting metrics like *who's active and how much* and *time spent on the platform*. AI summarized their responses as feature requests. A critical designer notices that students preface these suggestions with disclaimers such as *Not hindering anyone's privacy should be a priority.* The tension between the question's premise and students' discomfort reveals that we may be asking the wrong question entirely; perhaps the issue is not how to track engagement, but whether instructors should track it at all.

**4. Value judgments:** Deciding which contradictions to privilege, whose needs to prioritize, what kind of user experience to create. Example from our data: Students present contradictory needs around private chat (Q3-Q4, section B). Some students say, "*it is the biggest cause of distraction*" and should be restricted to students sending personal messages to teachers only. Others note it is crucial for people with anxiety or fear of judgement who cannot speak aloud. AI identified this as a contradiction requiring resolution. But this is not an empirical question with a correct answer, it is an ethical design decision about whose needs to privilege: neurotypical students who want fewer distractions, or anxious students who need alternative participation channels. Designer(s) must decide whether to optimize for focus or accessibility, knowing either choice disadvantages someone.

**5. Domain-specific context:** Understanding that teachers do not encourage discussion is institutional critique or pedagogical choice, not a feature request. Example from our data: When discussing breakout rooms (Q4-Q5, section C), one student observed: *"Teachers do not encourage separate group discussions and mostly focus on offering lectures and covering topics."* AI tagged this as a feature request to make breakout rooms more prominent and accessible. A designer with educational technology experience recognizes this as institutional critique about pedagogical culture. The problem is not that breakout rooms are hard to find; it is that lecture-based teaching norms do not value collaborative learning. Making the button bigger would not change teaching practices. This requires faculty development, curricular reform, or institutional policy to think of interventions outside the technology scope.

**6. Connecting the Dots:** Connecting micro-patterns to macro-themes. Example from our data: Across multiple questions, students exhibit meta-level self-reflection about their own suggestions. In Q5 (section A): "*Do people not know which situation to use and what layout?"* In Q4 (section B, about private chat): *"Maybe it is not really about having this feature or not but about having something else."* In section D, Q6 about fun elements: *"It should not be only about finishing a task or a topic but also making sure that everyone is comfortable enough."* AI treated these as individual comments within their respective questions. A designer recognizes a in between lines as students are questioning the premise of our questions, revealing uncertainty about whether technology fixes can address pedagogical or social problems. This pattern suggests the research itself needs to expand beyond feature requests to investigate classroom dynamics, power relations, and learning culture.

## FRAMEWORK TOWARDS AUGMENTED SENSEMAKING

Based on our comparative analysis, we propose a framework (work in progress) of **augmented sensemaking** that positions AI as an instrument for amplifying human judgment rather than replacing it. This framework has three core principles:

### 1. Distributed Labor between AI and Designer

AI should handle mechanical tasks (initial coding, frequency counting, quote extraction) while designers retain authority over interpretive work (prioritization, contradiction resolution, reading silences). This division of labor mirrors Kittur et al. [10] findings about crowd-powered qualitative coding: mechanical decomposition can be distributed, but synthesis requires expert judgment.

*Practical implementation:* Use AI for initial coding with predefined categories but make sure it requires human review of all ambiguous cases and outliers. Generate AI summaries alongside preserved raw quotes so that designers can verify interpretations against original context.

### 2. Contradiction as Resource and not Errors

Rather than asking AI to resolve contradictions through averaging or consensus, prompt it to identify and highlight tensions. In our data, the contradiction between wanting minimal interfaces and requesting more features is not a problem to solve, it is a design space to explore. As Suchman [8] argues, contradictions often mark sites where different stakeholder needs diverge, making them valuable for understanding power dynamics.

*Practical implementation:* Explicitly prompt AI to find conflicting responses and contradictory themes. Use these contradictions as discussion prompts in design workshops, forcing teams to make explicit value judgments about which user needs to privilege.

### 3. AI Outputs as Provocations and not Verdicts

Treat AI-generated analyses as provocations that spark designer reflection rather than definitive findings. When AI categorizes a response, ask: What perspective makes this categorization sensible? What alternative readings are being foreclosed? This approach aligns with critical data studies [11], which emphasize that all data analysis involves situated interpretation.

*Practical implementation:* After generating AI themes, conduct a challenge round where designers propose alternative interpretations. Compare AI categories with human-generated codes to identify divergences worth investigating. Use disagreement as a site for methodological reflection.

## IMPLICATIONS FOR DESIGN PRACTICE

Our findings have several implications for design practitioners integrating AI into qualitative research workflows:

### Resist the Efficiency

The primary appeal of AI sensemaking tools is speed, analyzing in minutes what might take days by hand. However, sensemaking is not just about reaching conclusions, it is about dwelling with ambiguity, noticing unexpected patterns, and developing tacit knowledge about the user in enquiry [12]. Rushing through analysis to deployment may produce designs that might be efficient but lack insight.

### Make Subjectivity Intentional

AI tools often promise to reduce bias by applying consistent criteria but eventually makes a formalized one. However, the choice of criteria is itself subjective. Rather than trying to eliminate bias, designers should make their interpretive choices explicit and accountable. When you decide to prioritize accessibility over efficiency, or to interpret uncertainty as meaningful rather than noise, document these judgments as design rationale.

### Preserve Raw Data Access

Always maintain access to original user responses alongside AI-generated summaries. Our analysis revealed multiple instances where context was essential to interpretation, a student in my view signaling personal opinion rather than fact, another's question marks indicating uncertainty. Summaries inevitably lose this texture (variety).

### Develop Critical AI Literacy

Design teams need training not just in using AI tools, but in critically evaluating their outputs. This includes understanding what AI tools can and cannot do (pattern matching versus understanding), recognizing when AI perpetuates problematic assumptions (our engagement tracking example), and knowing when to override AI suggestions based on domain expertise.

## CONCLUSION

The question is not whether AI can help with sensemaking, it demonstrably can. The question is *what kind of understanding we produce when we outsource interpretation to these AI.* Design is not just about identifying patterns; it is about making judgments in the web of data, contradictory user needs, and competing values. Our analysis of 20 student responses about video conferencing reveals that AI sensemaking tools are most valuable when they remain *instruments*, not *authorities*. They should amplify the designer's capacity to have attention to details, flagging low frequency user data as a major pain point or flaw, ensuring comprehensiveness, providing rapid initial organization, without indicating a direction, ethics, and situated knowledge that make design research meaningful. The goal of augmented sensemaking is not to eliminate human bias or subjectivity, these are essential to design judgment. Rather, it is to make that subjectivity *intentional*, *reflective*, and *accountable*. When a designer decides to prioritize accessibility over efficiency, to read uncertainty as meaningful, or to interpret silence as significant, that judgment should be explicit and justifiable, not washed away by AI objectivity. As AI tools become increasingly integrated into design workflows, we must resist the seduction of efficiency that comes at the cost of insight. The interpretive labor of sensemaking, dwelling with ambiguity, noticing what is missing, questioning our own assumptions, is precisely what makes design research valuable. AI can augment this labor, but it cannot replace it without fundamentally transforming what it means to understand users (which it may never achieve). At the workshop, we hope to explore with participants: What forms of sensemaking are worth preserving from automation? How do we design AI tools that respect, rather than erase, the interpretive agency of designers? What practices can ensure that AI efficiency enhances rather than diminishes the quality of design understanding?

## REFERENCES

**1.** Russell, D. M., Stefik, M. J., Pirolli, P., and Card, S. K. 1993. The cost structure of sensemaking. In Proceedings of INTERACT'93 and CHI'93 (Amsterdam, The Netherlands, 24-29 April 1993), ACM, 269-276. https://doi.org/10.1145/169059.169073

**2.** Muller, M., Lange, I., Wang, D., Piorkowski, D., Tsay, J., Liao, Q. V., Dugan, C., and Erickson, T. 2019. How data science workers work with data: Discovery, capture, curation, design, creation. In Proceedings of the 2019 CHI Conference on Human Factors in Computing Systems (Glasgow, Scotland, 4-9 May 2019), Paper 127, 1-15. https://doi.org/10.1145/3290605.3300356

**3.** Hook, K., & Lowgren, J. 2012. Strong concepts: Intermediate-level knowledge in interaction design research. ACM Trans. Comput.-Hum. Interact. 19, 3, Article 23 (August 2012), 18 pages. https://doi.org/10.1145/2362364.2362371

**4.** Dang, A., Goyal, S., Muller, M., Liao, Q. V., & Candello, H. 2024. Generative AI for qualitative research: Emerging practices and ethical considerations. In Extended Abstracts of the 2024 CHI Conference on Human Factors in Computing Systems (Honolulu, HI, 11-16 May 2024), ACM. https://doi.org/10.1145/3613905.3636294

**5.** Gaver, W. 2012. What should we expect from research through design? In Proceedings of the SIGCHI Conference on Human Factors in Computing Systems (Austin, Texas, 5-10 May 2012), ACM, 937-946. https://doi.org/10.1145/2207676.2208538

**6.** Jansen, Y., Dragicevic, P., Isenberg, P., Alexander, J., Karnik, A., Kildal, J., Subramanian, S., and Hornbaek, K. 2015. Opportunities and challenges for data physicalization. In Proceedings of the 33rd Annual ACM Conference on Human Factors in Computing Systems

(Seoul, Republic of Korea, 18-23 April 2015), ACM, 3227-3236. https://doi.org/10.1145/2702123.2702592

**7.** Koesten, L., Gregory, K., Groth, P., and Simperl, E. 2021. Talking datasets - understanding data sensemaking behaviours. Int. J. Hum.-Comput. Stud. 146 (February 2021), 102562. https://doi.org/10.1016/j.ijhcs.2020.102562

**8.** Suchman, L. A. 2007. *Human-Machine Reconfigurations: Plans and Situated Actions* (2nd ed.). Cambridge University Press. ISBN 978-0-521-67044-7

**9.** Shklovski, I., Mainwaring, S. D., Skúladóttir, H. H., & Borgthorsson, H. 2014. Leakiness and creepiness in app space: Perceptions of privacy and mobile app use. In Proceedings of the SIGCHI Conference on Human Factors in Computing Systems (Toronto, Canada, 26-April 01, 2014), ACM, 2347-2356. https://doi.org/10.1145/2556288.2556421

**10.** Kittur, A., Peters, A. M., Dirla, A., Thaker, T., & Dow, S. P. 2013. The crowd is a collaborative qualitative coding machine. In Proceedings of the 2013 Conference on Computer Supported Cooperative Work (San Antonio, Texas, 23-27 February 2013), ACM, 1691-1699. (Note: Common citation as CSCW; DOI https://doi.org/10.1145/2441776.2441966

**11.** Gitelman, L. (Ed.). (2013). *Raw data is an oxymoron*. MIT Press. ISBN 978-0-262-51977-6

**12.** Schon, D. A. 1983. *The Reflective Practitioner: How Professionals Think in Action.* Basic Books. ISBN 978-0-465-06878-4

**13.** Zimmerman, J., Forlizzi, J., & Evenson, S. 2007. Research through design as a method for interaction design research in HCI. In Proceedings of the SIGCHI Conference on Human Factors in Computing Systems (San Jose, California, 28 April - 03 May 2007), ACM, 493-502. https://doi.org/10.1145/1240624.1240704